\documentclass{llncs}
\usepackage{eepic,latexsym,amssymb}
\usepackage[all]{xy}
\usepackage{fancybox,epic}

\usepackage{listings}
\def\dep#1{\langle{#1}\rangle}

\newcommand{\selatpar}[1]{A_{R#1}}
\newcommand{\selat}{A_{R}}
\newcommand{\select}{\cR}

\newcommand{\cR}{{\cal R}}

\newcommand{\tuple}[2]{#1:#2}

\newcommand{\ciao}{{\tt Ciao}}
\newcommand{\ciaopp}{{\tt CiaoPP}}

\newcommand{\mycomment}[1]{*** \textbf{#1} ***}

\long\def\comment#1{}

\newcommand{\CP}{\mbox{\it CP}}

\newcommand{\AP}{\mbox{\it AP}}

\newcommand{\sizeinfigure}{\small} 

\usepackage{algorithm}
\usepackage[noend]{algpseudocode}

\newcommand{\arestrict}{\mbox{\sf Arestrict}}
\newcommand{\aextend}{\mbox{\sf Aextend}}
\newcommand{\aunif}{\mbox{\sf Aunif}}
\newcommand{\aconj}{\mbox{\sf Aconj}}
\newcommand{\alub}{\mbox{\sf Alub}}
\newcommand{\acalltoentry}{\mbox{\sf Atranslate}}

\newcommand{\at}{\mbox{{$\cal AT$}}}
\newcommand{\dt}{\mbox{{$\cal DT$}}}
\newcommand{\gt}{\mbox{{$\cal GT$}}}
\newcommand{\st}{\mbox{{$\cal ST$}}}
\newcommand{\emptytable}{\mbox{\sf Create\_Table}}
\newcommand{\store}{\mbox{\sf Insert}}
\newcommand{\isin}{\mbox{\sf IsIn}}
\newcommand{\consult}{\mbox{\sf Look\_up}}

\newcommand{\newpred}{\mbox{\sf new\_filter}}

\newcommand{\ren}{\mbox{\sf ren}}
\newcommand{\getbody}{\mbox{\sf get\_body}}

\usepackage{listings}
\newcommand{\aabstraction}{\mbox{\it AGeneralize}}
\newcommand{\aunfold}{\mbox{\it AUnfold}}
\newcommand{\abstraction}{\mbox{\it Generalize}}
\newcommand{\unfold}{\mbox{\it Unfold}}

\newcommand{\widencall}{\mbox{\it Widen\_Call}}

\newcommand{\short}[1]{}
\newcommand{\secbeg}{\vspace*{0mm}}
\newcommand{\secend}{\vspace*{0mm}}

\title{A Generic Framework for the Analysis and Specialization of
  Logic Programs}
\author{Germ\'{a}n Puebla\inst{1} \and Elvira Albert\inst{2}
 \and Manuel Hermenegildo\inst{1,3}
}

\institute{
School of Computer Science, Technical U.~of Madrid, 
\email{\{german,herme\}@fi.upm.es}
\and
School of Computer Science, Complutense U.~of  Madrid,
\email{elvira@sip.ucm.es}
\and
Depts. of Comp.~Sci.~and El.~and Comp.~Eng., U. of  New Mexico,
\email{herme@unm.edu}
} 

\begin{document}
\maketitle

 \pagestyle{myheadings}  \markboth{WLPE 2005}{Generic Framework for Analysis and Specialization
 of Logic Programs} 


\begin{abstract}
   

The relationship between abstract interpretation and partial deduction
has received considerable attention and (partial) integrations have
been proposed starting from both the partial deduction and abstract
interpretation perspectives.
In this work we present what we argue is the first fully described
generic algorithm for efficient and precise integration of abstract
interpretation and partial deduction.
%
Taking as starting point state-of-the-art algorithms for
context-sensitive, polyvariant abstract interpretation and (abstract)
partial deduction, we present an algorithm which combines the best of
both worlds. Key ingredients include the accurate success propagation
inherent to 
abstract interpretation and the powerful
program transformations achievable by partial deduction.
In our algorithm, the calls which appear 
in the analysis graph are not analyzed w.r.t.~the original definition
of the procedure
but w.r.t.
\emph{specialized definitions} of these procedures.  Such specialized
definitions are obtained by applying
both unfolding and abstract executability.
Our framework 
is parametric w.r.t.~different control strategies and abstract
domains. Different combinations of such parameters correspond to
existing algorithms for program analysis and specialization.
Simultaneously, our approach opens the door to the efficient
computation of strictly more precise results than those achievable by
each of the individual techniques.
\short{On one hand, the algorithm computes a precise
description of the behaviour of the (residual) program.
On the other hand, the specialized definitions computed during the
execution of the algorithm provide a highly specialized partial
evaluation of the program.
%
} 
The algorithm is now one of the key components of the \ciaopp\ 
analysis and specialization system.

\end{abstract}

\vspace{-0.5cm}
\section{Introduction and Motivation}
\secend
 \label{sec:motivation}
 
\short{
 \input motivation
 
\section{Running Example}
\label{sec:running-example}
} 

The relationship between abstract interpretation~\cite{Cousot77-short}
and partial evaluation~\cite{pevalbook93} has received considerable
attention (see for example
\short{pe-in-plai-ws-short,}
\cite{gallag88,gallagher:wsa92,Consel-Koo:toplas93,LeuschelDeSchreye:PLILP96-short,jones-pe-ai97,leuschel-jicslp98,spec-jlp-short,pe-in-plai-pepm99-short,Gallagher-Peralta-HOSC,leuschel-gruner-loprstr2001-short,Cousot02-short,spec-inv-pepm03-short,leuschel04:Toplas}
and their references).
In order to motivate and illustrate our proposal for an integration of
abstract interpretation and partial evaluation, we use the running
example of Fig.~\ref{lst:running-ex}. It is a simple \ciao \short{
  \ciao~\cite{ciao-manual-1.10-short}
} 
program which uses Peano's arithmetic.\footnote{Rules are written with a unique
subscript attached to the head atom (the rule number), and a dual
subscript (rule number, body position) attached to each body literal.
We sometimes use this notation for denoting calls to atoms as well.}
We use the \ciao\ assertion
language 
in order to provide
precise descriptions on the initial call patterns. In our case, the
\texttt{entry} declaration is used to inform that all calls to the
only exported predicate (i.e., \texttt{main/2}) will always be of the
form $\tt \leftarrow main(s(s(s(L))),R)$ with $\tt L$ ground and $\tt
R$ a variable.
The predicate \texttt{main/2} performs two calls to
predicate \texttt{formula/2}, which contains mode tests \texttt{ground(X)}
and \texttt{var(W)} on its input arguments. A call
\texttt{formula(X,W)} returns $W=(X-2)\times 2$. 
Predicate \texttt{two/1} returns the natural number 2 in Peano's
arithmetic.
A call \texttt{minus(A,B,C)} returns $C=B-A$. However, if the result
becomes a negative number, $C$ is left as a free variable. This
indicates that the result is not valid.
In turn, a call \texttt{twice(A,B)} returns $B=A\times 2$. 
Prior to computing the result, this predicate checks whether $A$ is
valid, i.e., not a variable, and simply returns a variable otherwise.

 \begin{figure}[t]
{\footnotesize
\tt
:- module(\_,[main/1],[assertions]).\\
:- entry main(s(s(s(L))),R) : (ground(L),var(R)).\\[0.6ex]
main$_{1}$(X,X2):- formula$_{1,1}$(X,X1), formula$_{1,2}$(X1,X2).\\[0.6ex]
formula$_{2}$(X,W):- ground$_{2,1}$(X),var$_{2,2}$(W),two$_{2,3}$(T),minus$_{2,4}$(T,X,X2),twice$_{2,5}$(X2,W).\\[0.6ex]
two$_{3}$(s(s(0))).\\[0.6ex]
minus$_{4}$(0,X,X).\\
minus$_{5}$(s(X),s(Y),R):- minus$_{5,1}$(X,Y,R).\\
minus$_{6}$(s(\_X),0,\_R).\\[0.6ex]
twice$_{7}$(X,\_Y):- var$_{7,1}$(X).\\
twice$_{8}$(X,Y):- ground$_{8,1}$(X), tw$_{8,2}$(X,Y).\\[0.6ex]
tw$_{9}$(0,0).\\
tw$_{10}$(s(X),s(s(NX))):- tw$_{10,1}$(X,NX).
}
\caption{Running Example}
\label{lst:running-ex}
\end{figure}


By observing the behaviour of the program it can be seen that for
initial queries satisfying the \texttt{entry} declaration, all calls
to the tests \texttt{ground$_{2,1}$(X)} and \texttt{var$_{2,2}$(W)}
will definitely succeed, even if we do not know the concrete values of
variable $\tt L$ at compile time.  Also, the calls to
\texttt{ground$_{8,1}$(X)} will succeed, while the calls to
\texttt{var$_{7,1}$(X)} will fail. This shows the benefits of (1)
\emph{exploiting abstract information in order to abstractly execute
  certain atoms, which in turn may allow unfolding of other atoms}.
However, the use of an abstract domain which captures groundness and
freeness information will in general not be sufficient to determine
that in the second execution of \texttt{formula/2} the tests
\texttt{ground$_{2,1}$(X)} and \texttt{var$_{2,2}$(W)} will also
succeed. The reason is that, on success of
\texttt{minus$_{2,4}$(T,X,X2)}, $\tt X2$ cannot be guaranteed to be
ground since \texttt{minus$_6$/3} succeeds with a
free variable on its third argument position. It can be observed,
however, that for all calls to \texttt{minus/3} in executions
described by the \texttt{entry} declaration, such third clause for
\texttt{minus/3} is useless. It will never contribute to a success of
\texttt{minus/3} since such predicate is always called with a value
greater than zero on
its second argument.
Unfolding can make this explicit by fully unfolding calls to
\texttt{minus/3} since they are sufficiently instantiated (and as a
result the ``dangerous'' third clause is disregarded).  It allows
concluding that in our particular context, all calls to
\texttt{minus/3} succeed with a ground third argument.
This shows the importance of (2) \emph{performing unfolding steps in
  order to prune away useless branches, which will result in improved
  success information}. By the time execution reaches
\texttt{twice$_{2,5}$(X2,W)}, we hopefully know that $\tt X2$ is
ground.  In order to determine that, upon success of
\texttt{twice$_{2,5}$(X2,W)} (and thus on success of
\texttt{formula$_{1,1}$(X,W)}), $\tt W$ is ground, we need to perform
a fixpoint computation. Since, for example, the success substitution
for \texttt{formula$_{1,1}$(X,X1)} is indeed the call substitution for
\texttt{formula$_{1,2}$(X1,X2)}, the success of the second test
\texttt{ground$_{2,1}$(X)} (i.e., the one reachable from
\texttt{formula$_{1,2}$(X1,X2)}) cannot be established unless we
propagate success substitutions. This illustrates the importance of
(3) \emph{propagating (abstract) success information, performing
  fixpoint computations when needed, which simultaneously will result
  in an improved unfolding}. Finally, whenever we call
\texttt{formula(X,W)}, $\tt W$ is a variable, a property which cannot be
captured if we restrict ourselves to downwards-closed domains. This
indicates (4) \emph{the usefulness of having information on non
  \emph{downwards-closed} properties}.

Throughout the paper we show that the framework we propose is able to
eliminate all calls to mode tests \texttt{ground/1} and
\texttt{var/1}, and predicates \texttt{two/1} and \texttt{minus/3} are
both fully unfolded and no longer appear in the residual code.
We have used \emph{sharing--freeness} as abstract domain instead of
one based on, say regular types, for two reasons.\footnote{The values
  for the rest of parameters are: \aabstraction\ and \aunfold\ rules
  based on \emph{homeomorphic
    embedding}~\cite{LeuschelBruynooghe:TPLP02}, and the identity
  function as \widencall\ function.}  First, to illustrate how
non-downwards closed information, including freeness and definite
independence, can be correctly exploited by our algorithm
in order to 
optimize the
program, and second, to show how unfolding can be of great use in
order to improve the accuracy of analyses apparently unrelated to
partial deduction, such as the classical \emph{sharing--freeness}.
\begin{example}\label{ex:sp}
  The results obtained by \ciaopp---which implements abstract
  interpretation with specialized definitions---are both the following
  specialized code and an accurate analysis for such
  program (rules are renamed using the prefix $\tt sp$).
%
  {\footnotesize
\begin{quote}\tt
sp\_main${}_1$(s(s(s(0))),0).\\
sp\_main${}_2$(s(s(s(s(B)))),A) :- sp\_tw${}_{2,1}$(B,C),
sp\_formula${}_{2,2}$(C,A).\\[0.6ex]
sp\_tw${}_2$(0,0).\\
sp\_tw${}_3$(s(A),s(s(B))) :-  sp\_tw${}_{3,1}$(A,B).\\[0.6ex]
sp\_formula${}_4$(0,s(s(s(s(0))))).\\
sp\_formula${}_5$(s(A),s(s(s(s(s(s(B))))))) :-  sp\_tw${}_{5,1}$(A,B).
\end{quote}
}
\noindent
In this case, the success information for \texttt{sp\_main(X,X2)}
guarantees that $\tt X2$ is definitely ground on success. Note that
this is equivalent to proving $\forall X\geq 3,~main(X,X2)\rightarrow
X2\geq 0$. Furthermore, our system is able to get to that conclusion
even if the \texttt{entry} only informs about $\tt X$ being any
possible ground term and $\tt X2$ a free variable.
\end{example}
The above results cannot be achieved unless all four points mentioned
before are available in a program analysis/specialization system.
For example, if we use traditional partial
deduction~\cite{Lloyd91,gallagher:pepm93-short} (PD) with the corresponding
\abstraction\ and \unfold\ rules followed by abstract interpretation
and \emph{abstract specialization} as described
in~\cite{spec-jlp-short,spec-inv-pepm03-short} we only obtain a comparable program
after four iterations of the: ``PD + abstract
interpretation + abstract specialization'' cycle.
If we keep on adding more calls to {\tt formula}, every time more
iterations are necessary to obtain results comparable to ours. This
shows the importance of achieving an algorithm which is able to
\emph{interleave} PD, with abstract interpretation, extended with abstract
specialization, in order to communicate
the accuracy gains achieved from one to the other as soon as possible.
In any case, iterating over ``PD + analysis'' is not a
good idea from the efficiency point of view. Also, sometimes partially
evaluating a partially evaluated program can degrade the quality of
the residual program.

The remaining of the paper is organized as follows.
Section~\ref{sec:prel-notat} recalls some preliminary concepts. In
Sect.~\ref{sec:unfold-with-abstr}, we present abstract unfolding which
already integrates abstract executability.
Section~\ref{sec:spec-defin} introduces our notion of specialized
definition and embeds it within an abstract partial deducer. In
Sect.~\ref{sec:effic-prop-abstr}, we propose our scheme for abstract
interpretation with specialized definitions.
Finally, Sect.~\ref{sec:disc-relat-work}
compares to related work and Sect.~\ref{sec:benefits-or-our}
concludes.

\secbeg
\section{Preliminaries} 
\secend
\label{sec:prel-notat}

Very briefly (see for example~\cite{Lloyd87} for details), an
\emph{atom} $A$ is a syntactic construction of the form
$p(t_1,\ldots,t_n)$, where $p/n$, with $n\geq 0$, is a predicate
symbol and $t_1,\ldots,t_n$ are terms. 
A \emph{clause} is of the form $H\leftarrow B$ where its head $H$ is
an atom and its body $B$ is a conjunction of atoms. A \emph{definite
  program} is a finite set of clauses. A \emph{goal} (or query) is a
conjunction of atoms.

Let $G$ be a goal of the form
  $\leftarrow A_1,\ldots,\selat,\ldots,A_k$, $k\geq 1$. 
The concept of
  \emph{computation rule}, denoted by $\select$,  is used
to select an atom within a goal for its evaluation.
%
 %
%
%
The operational semantics of definite programs is based on
derivations~\cite{Lloyd87}.
%
Let $C=H \leftarrow B_1,\ldots,B_m$ be a renamed apart clause in $P$
such that $\exists \theta=mgu(\selat,H)$. Then $\leftarrow
\theta(A_1,\ldots,\selatpar{-1},B_1,\ldots,B_m,\selatpar{+1},\ldots,A_k)$
is \emph{derived} from $G$ and $C$ via $\select$.
%
As customary, given a program $P$ and a goal $G$, an \emph{SLD
  derivation} for $P\cup\{G\}$ consists of a possibly infinite
sequence $G=G_0, G_1, G_2,\ldots$ of goals, a sequence $C_1,C_2,\ldots$
of properly renamed apart clauses of $P$, and a sequence
$\theta_1,\theta_2,\ldots$ of mgus such that each $G_{i+1}$ is derived
from $G_i$ and $C_{i+1}$ using $\theta_{i+1}$.
A derivation step can be non-deterministic when $\selat$ unifies with
several clauses in $P$, giving rise to
several possible SLD derivations for a given goal.  Such SLD
derivations can be organized in \emph{SLD trees}.
A finite derivation $G=G_0, G_1, G_2,\ldots, G_n$ is called
\emph{successful} if $G_n$ is empty.  In that case
$\theta=\theta_1\theta_2\ldots\theta_n$ is called the computed answer
for goal $G$.  Such a derivation is called \emph{failed} if it is not
possible to perform a derivation step with $G_n$.
%
Given an atom $A$, an \emph{unfolding
  rule}~\cite{Lloyd91,gallagher:pepm93-short} computes a set of finite SLD
derivations $D_1,\ldots,D_n$ (i.e., a possibly incomplete SLD tree) of
the form $D_i=A,\ldots,G_i$ with computed answer substitution
$\theta_i$ for $i=1,\ldots,n$ whose associated \emph{resultants} (or
residual rules) are $\theta_i(A) \leftarrow G_i$. 

The following standard operations are used in the paper to handle
keyed-tables: \emptytable$(T)$ initializes a table $T$.
\store$(T,\mathit{Key},\mathit{Info})$ adds $\mathit{Info}$ associated
to $\mathit{Key}$ to $T$ and deletes previous information associated
to $\mathit{Key}$, if any.  \isin$(T,\mathit{Key})$ returns true iff
$\mathit{Key}$ is currently stored in the table. Finally,
\consult$(T,\mathit{Key})$ returns the information associated to
$\mathit{Key}$ in $T$.  For simplicity, we sometimes consider tables
as sets and we use the notation $(\mathit{Key}\leadsto
\mathit{Info})\in T$ to denote that there is an entry in the table T
with the corresponding $\mathit{Key}$ and associated $\mathit{Info}$.

\secbeg
\subsection{Abstract Interpretation}
\secend

Abstract interpretation \cite{Cousot77-short} provides a general
formal framework for computing safe approximations of programs
behaviour.
%
Programs are interpreted using values in an \emph{abstract domain}
($D_\alpha$) instead of the \emph{concrete domain} ($D$).
%
The set of all possible abstract  values which represents
$D_\alpha$ is usually a complete lattice or cpo which is ascending
chain finite.
%
Values in the abstract domain $\langle D_\alpha, \sqsubseteq \rangle$
and sets of values in the concrete domain $\langle 2^D, \subseteq
\rangle$ are related via a pair of monotonic mappings $\langle \alpha,
\gamma \rangle$: the {\em abstraction} function $\alpha:
2^D\rightarrow D_\alpha$ which assigns to each (possibly infinite) set
of concrete values an abstract value, and the {\em concretization}
function $\gamma:D_\alpha\rightarrow 2^D$ which assigns to each
abstract value the (possibly infinite) set of concrete values it
represents.
\short{such that
$\begin{array}{c}
 \forall x\in 2^D:~ \gamma(\alpha(x)) \supseteq x \mbox{~~~and~~~}
 \forall y\in D_\alpha:~ \alpha(\gamma(y)) = y .
\end{array}$
%
In general $\sqsubseteq$ is induced by $\subseteq$ and $\alpha$.
} 
The operations on the abstract domain $D_{\alpha}$ that we will use in
our algorithms are:
\begin{itemize}
\item \arestrict$(\lambda,E)$ performs the abstract restriction (or
  projection) of a substitution $\lambda$ to the set of variables in
  the expression $E$, denoted $vars(E)$;
\item \aextend$(\lambda,E)$ extends the substitution $\lambda$ to the
  variables in the set $vars(E)$;
\item \aunif$(t_1,t_2,\lambda)$ obtains the description which results
  from adding the abstraction of the unification $t_1 = t_2$ to the
  substitution $\lambda$;
\item \aconj$(\lambda_1,\lambda_2)$ performs the abstract conjunction
  ($\sqcap$) of two substitutions;
\item \alub$(\lambda_1,\lambda_2)$ performs the abstract
  disjunction ($\sqcup$) of two substitutions.
\end{itemize}
%
In our algorithms we also use \acalltoentry$(A:CP,H\leftarrow B)$
which adapts and projects the information in an abstract atom $A:CP$
to the variables in the clause $C=H\leftarrow B$.  An \emph{abstract atom} of the form
$G:CP$ is a concrete atom $G$ which comes equipped with an
\emph{abstract substitution} $CP$ which is defined over $vars(G)$ and
provides additional information on the context in which the atom will
be executed at run-time.  \acalltoentry\ can
be defined in terms of the operations above as:
\acalltoentry$(A:CP,H\leftarrow B)$ =
$\arestrict$(\aunif$(A,H,\aextend(CP, C)),C)$.
Finally, the most general substitution is represented as $\top$,
and the least general (empty) substitution as $\bot$.

\secbeg
\section{Unfolding with Abstract Substitutions}\label{sec:unfold-with-abstr}
\secend

We now present an extension of SLD semantics which exploits abstract
information. This will provide the means to overcome difficulties (1)
and (2) introduced in Section~\ref{sec:motivation}.
\short{The objective is to improve the quality of specialized rules
  (or resultants) by both unfolding and exploiting abstract call
  information.
We start by recalling some preliminary notions on partial evaluation.
} 
 The extended semantics handles abstract goals of the form
$G:CP$, i.e., a concrete goal $G$ comes equipped with an
abstract substitution $CP$ defined over $vars(G)$.
%
The first rule 
corresponds to a derivation step.
\begin{definition}[derivation step]
 \label{def:der-step-abs}
 Let $G:CP$ be an abstract goal where $G=\leftarrow
 A_1,\ldots,\selat,\ldots,A_k$. Let $\select$
be a computation rule and let $\select(G)=$$\selat$. 
Let $C=H \leftarrow B_1,\ldots,B_m$ be a renamed apart clause in $P$.
Then the abstract goal $G':CP'$ is \emph{derived} from $G:CP$ and $C$
via $\select$ if $\exists \theta=mgu(\selat,H)~\wedge$ $CP_u \neq
\bot$, where:
\begin{eqnarray*}
CP_u =\aunif(\selat,H\theta,\aextend(CP,C\theta))\\
G'=\theta(A_1,\ldots,\selatpar{-1},B_1,\ldots,B_m,\selatpar{+1},\ldots,A_k)\\
CP'=\arestrict(CP_u,vars(G'))
\end{eqnarray*}
\end{definition}
%
%
An important difference between the above definition and the standard
derivation step
is that the use of abstract (call) substitutions allows imposing further
conditions for performing derivation steps, in particular, $CP_u$
cannot be $\bot$.
This is because if $CP\neq\bot$ and $CP_u=\bot$ then the head of the
clause $C$ is incompatible with $CP$ and the unification $\selat=H$ will
definitely fail at run-time.  \short{Thus, a derivation step is not
  allowed if $CP_u=\bot$ which may result in a failed derivation if
  there are not other clauses to resolve the selected atom $\selat$.}
Thus, abstract information allows us to remove useless clauses from
the residual program.
This produces more efficient resultants and increases the accuracy of
analysis for the residual code.
\begin{example}\label{ex:derive}
  Consider the abstract atom $\tt
  \tuple{formula(s^4(X),X2)}{\{X/G,X2/V\}}$, which appears in the
  analysis of our running example (c.f. Fig.~\ref{fig:aog}).  We
  abbreviate as $\tt s^n(X)$ the successive application of $\tt n$
  functors $\tt s$ to variable $\tt X$. The notation $\tt X/G$ (resp.
  $\tt X/V$) indicates that variable $\tt X$ is ground (resp. a free
  variable).  After applying a derivation step, we obtain the derived
  abstract goal:\\[0.6ex]
  {\small $\tt
    \tuple{ground(s^4(X)),var(X2),two(T),minus(T,s^4(X),X2'),twice(X2',X2)}{\{X/G,X2/V,T/V,X2'/V\}}
    $}\\[0.6ex]
  where the abstract description has been extended with updated
  information about the freeness of the newly introduced variables. In
  particular, both {\tt T} and {\tt X2'} are {\tt V}.
\end{example}


\comment{
Regarding rule $\mathit{Aexec}$, though it may seem of narrow
applicability, in fact for many builtin procedures such as those that
check basic types or which inspect the structure of data, even these
simple optimizations are indeed very relevant.  Two non-trivial
examples of this are their application to simplifying independence
tests in program parallelization~\cite{spec-jlp-short} and the optimization
of delay conditions in logic programs with dynamic procedure call
scheduling order~\cite{spec-dyn-sch-iclp97}. 

the incomplete SLD trees computed when compared to those achievable
when traditional SLD resolution is used. This consequently improves
the quality of the unfolding process and thus we generate
better resultants. 
}


The second rule we present makes use of the availability of abstract
substitutions to perform \emph{abstract
  executability}~\cite{spec-jlp-short} during resolution.  This allows
replacing some atoms with simpler ones, and, in particular, with the
predefined atoms {\em true} and {\em false}, provided certain
conditions hold.
We assume the existence of a predefined \emph{abstract executability
  table} which contains entries of the form $T:CP \leadsto T'$ which
specify the behaviour of external procedures: builtins, libraries, and
other user modules.  For instance, for predicate $\tt ground$ contains
the information $\tt ground(X):\{X/G\} \leadsto true$. For $\tt var$,
it contains $\tt var(X):\{X/V\} \leadsto true$.\footnote{In \ciaopp\ we
  use assertions to express such
  information in a domain-independent manner.}
\short{
The idea is that atoms of the form $A:CP$ which are produced during
unfolding and which are described by $T:CP_T$ such that $A=\theta(T)$
and $A:CP$ is described by $T:CP_T$ can be safely replaced by
$\theta(T')$. 
} 

\begin{definition}[abstract execution]
 \label{def:abs-ex-true}
 Let $G:CP$ be an abstract goal where $G=\leftarrow
 A_1,\ldots,\selat,\ldots,A_k$. Let $\select$ be a computation rule
 and let $\select(G)=$$\selat$.
 Let $(T:CP_T \leadsto T')$ be a renamed apart entry in the abstract
 executability table.  Then, the goal $G':CP'$ is \emph{abstractly
   executed} from $G:CP$ and $(T:CP_T \leadsto T')$ via $\select$ if
 $\selat=\theta(T)$ and $CP_A\sqsubseteq CP_T$, where
\begin{eqnarray*}
G' =A_1,\ldots,\selatpar{-1},\theta(T'),\selatpar{+1},\ldots,A_k\\
CP'= \arestrict(CP,G')\\ 
CP_A = \acalltoentry(A_R:CP,T\leftarrow true)
 \end{eqnarray*}
\end{definition}
\begin{example}\label{ex:aexec}
  From the derived goal in Ex.~\ref{ex:derive}, we can apply twice the
  above rule to abstractly execute the calls to {\tt ground} and {\tt
    var} and obtain:
\begin{center} $\tt
  \tuple{two(T),minus(T,s^4(X),X2'),twice(X2',X2)}{\{X/G,X2/V,T/V,X2'/V\}}
$
\end{center}
since both calls succeed by using the abstract executability table
described above and the information in the abstract substitution.
\end{example}
\begin{definition}[\aunfold]\label{def:aunfold}
  Let $\tuple{A}{CP}$ be an abstract atom and $P$ a program. We define
  $\aunfold(P,\tuple{A}{CP})$ as the set of \emph{resultants}
  associated to a finite (possibly incomplete) SLD tree computed by
  applying the rules of Definitions \ref{def:der-step-abs} and
  \ref{def:abs-ex-true} to $\tuple{A}{CP}$.
\end{definition}
%
%
The so-called \emph{local control} of PD ensures the termination of
the above process. For this purpose, the unfolding rule must
incorporate some mechanism to stop the construction of SLD derivations
(we refer to \cite{LeuschelBruynooghe:TPLP02} for details).
\begin{example}\label{ex:aunfold}
  Consider an unfolding rule $\aunfold$ based on homeomorphic
  embedding \cite{LeuschelBruynooghe:TPLP02} to ensure termination and
  the initial goal in Ex.~\ref{ex:derive}. The derivation continuing
  from Ex.~\ref{ex:aexec}  performs several additional derivation steps
  and abstract executions and branches (we do not include them due to space
  limitations and also because it is well understood).  The following
  resultants are obtained from the resulting tree:
\begin{verbatim}
      formula(s(s(s(s(0),s(s(s(s(0))))).
      formula(s(s(s(s(s(A))))),s(s(s(s(s(s(B))))))) :- tw(A,B)
\end{verbatim}
%
which will later be filtered and renamed resulting
in rules 4 and 5 of Ex.~\ref{ex:sp}.
\end{example}
%
%
%
%
It is important to note that SLD resolution with
abstract substitutions is not restricted to the left-to-right
computation rule.  However, it is well-known that non-leftmost
derivation steps can produce incorrect results if the goal contains
\emph{impure} atoms to the left of $\selat$.  More details can be
found, e.g., in~\cite{LeuschelEtAl:TPLP03}.
Also, abstract execution of non-leftmost atoms can be incorrect if the
abstract domain used captures properties which are not downwards
closed. A simple solution is to only allow leftmost abstract execution
for non-downwards closed domains (and non-leftmost for derivation
steps).

\secbeg
\section{Specialized Definitions}
\secend
\label{sec:spec-defin}

\begin{algorithm}[t]
\caption{Abstract Partial Deduction with Specialized Definitions}
\label{algo:pd-with-spec-defs}
\begin{algorithmic} [1]
\renewcommand{\baselinestretch}{0.5} \sizeinfigure
\Procedure{partial\_evaluation\_with\_spec\_defs}{$P,\{A_1:CP_1,\ldots,A_n:CP_n\}$}
\State \emptytable$(\gt)$; \emptytable$(\st)$
\For{$j = 1..n$}
\State  {\sc process\_call\_pattern}($A_j:CP_j$) 
\EndFor
\EndProcedure
\Statex
\Procedure{process\_call\_pattern}{$A:CP$}
\If {{\bf not }\isin$(\gt,A:CP)$}
\State $(A_1,A_1')\gets$ {\sc specialized\_definition}$(P,A:CP)$
\State $A_1:CP_1\gets$ \consult$(\gt,A:CP)$
\ForAll  {ren. apart clause $C_k=H_k \leftarrow B_{k}\in P$
  s.t. $H_k$ unifies with $A_1'$}
\State $CP_k$ $\gets$ \acalltoentry$(A_1':CP_1,C_k)$
\State {\sc process\_clause}$(CP_k,~B_{k})$
\EndFor
\EndIf
\EndProcedure
\Statex
\Procedure{process\_clause}{$CP,~B$}
\If {$B=(L,R)$}
\State $CP_L \gets \arestrict(CP,L)$
\State {\sc process\_call\_pattern}($L:CP_L$)  
\State {\sc process\_clause}$(CP,~R)$  
\Else 
\State $CP_B \gets \arestrict(CP,B)$
\State {\sc process\_call\_pattern}($B:CP_B$)
\EndIf
\EndProcedure
\Statex
\Function {specialized\_definition}{$P,A:CP$}
\State $A':CP'\gets \aabstraction(\st,A:CP)$
\State $\store(\gt,\tuple{A}{CP},\tuple{A'}{CP'})$
\If {\isin$(\st,A':CP')$}
\State $A''\gets $\consult$(\st,A':CP')$
\Else
\State $\mathit{Def}\gets \aunfold(P,\tuple{A'}{CP'})$
\State $A''\gets \newpred(A')$ 
\State $\store(\st,\tuple{A'}{CP'},A'')$
\State $\mathit{Def}'\gets \{(H' \leftarrow B) ~|~ (H\leftarrow B)\in \mathit{Def}\wedge
H'=\ren(H,\{A'/A''\})\}$
\State $P\gets P\bigcup \mathit{Def}'$
\EndIf
\State \Return $(A',A'')$
\EndFunction
\renewcommand{\baselinestretch}{1.0}\normalsize
\end{algorithmic}
\end{algorithm}

We now define an Abstract Partial Deduction (APD) algorithm whose
execution can later be \emph{interleaved} in a seamless way with a
state-of-the-art abstract interpreter.  For this it is essential that
the APD process can generate residual code \emph{online}.  Thus,
we need to produce a residual, specialized definition for a call
pattern as soon as we finish processing it. This will make it possible
for the analysis algorithm to have access to the improved definition.
This may increase the accuracy of the analyzer and addresses the
difficulty (2) described in Sect.~\ref{sec:motivation}.

Typically, PD is presented as an iterative process in which partial
evaluations are computed for the new generated atoms until they
\emph{cover} all calls which can appear in the execution of the
residual program. This is formally known as the \emph{closedness}
condition of PD \cite{Lloyd91}. In order to ensure termination of this
global process, the so-called \emph{global} control defines a
$\aabstraction$ operator (see \cite{LeuschelBruynooghe:TPLP02}) which
guarantees that the number of SLD trees computed is kept finite, i.e.,
it ensures the finiteness of the set of atoms for which partial
evaluation is produced.
%
However, the residual program is not generated until such iterative
process terminates.

Algorithm~\ref{algo:pd-with-spec-defs} presents an APD algorithm.
%
The main difference with standard algorithms is that the resultants
computed by \aunfold\ (L26) are added to the program during execution
of the algorithm (L30) rather than in a later code generation phase.
In order to avoid conflicts among the new clauses and the original
ones, clauses for specialized definitions are renamed with a fresh
predicate name (L29) prior to adding them to the program (L30).
The algorithm uses two global data structures. The
\emph{specialization table} contains entries of the form $A:\CP{}
\leadsto A'$.  The atom $A'$ provides the link with the clauses of the
specialized definition for $A:CP$.
The \emph{generalization table} stores the results of the
\aabstraction\ function and contains entries
  $A:CP\leadsto A':CP$ where $A':CP'$ is a generalization of $A:CP$.
\short{We refer by LX to line number X in the algorithm}

Computation is initiated by procedure {\sc
  partial\_evaluation\_with\_spec\_defs} (L1-4) which initializes the
tables and calls {\sc process\_call\_pattern} for each abstract atom
$A_i:CP_i$ in the initial set
to be partially evaluated.  
%
The task of {\sc process\_call\_pattern} is, if the atom has not been
processed yet (L6), to compute a specialized definition for it (L7)
and then process all clauses in its specialized definition by means of
calls to {\sc process\_clause} (L9-11).
%
Procedure {\sc process\_clause} 
traverses clause bodies, processing their corresponding atoms by means
of calls to {\sc process\_call\_pattern}, in a depth-first,
left-to-right fashion. The order in which pending call patterns
(atoms) are handled by the algorithm is usually not fixed in PD
algorithms. They are often all put together in a set.  The reason
for this presentation is to be as close as possible to our analysis
algorithm which enforces a depth-first, left-to-right traversal of
program clauses.
In this regard, the relevant point to note is that this algorithm does
not perform success propagation yet (difficulty 3).  
In L16, it becomes apparent that the atom(s) in $R$ will be analyzed
with the same call pattern $CP$ as $L$, which is to their left in
the clause.
This, on one hand, may clearly lead to substantial precision loss.
For instance, the abstract pattern $\tt formula(C,A):\{C/G,C/V\}$ which is
necessary to obtain the last two resultants of Ex.~\ref{ex:sp} cannot
be obtained with this algorithm.  In particular, we cannot infer the
groundness of $\tt C$ which, in turn, prevents us from abstractly
executing the next call to $\tt ground$ and, thus, from obtaining this
optimal specialization.
On the other hand, this lack of success propagation makes it difficult
or even impossible to work with non downwards closed domains, since
$CP$ may contain information which holds before execution of the
leftmost atom $L$ but which can be uncertain or even false after that.
In fact, in our example $CP$ contains the info $\tt C/V$, which
becomes false after execution of $\tt tw(B,C)$, since now $\tt C$ is
ground.
This problem is solved in the algorithm we present in the next
section, where analysis information flows from left to right, adding
more precise information and eliminating information which is no
longer safe or even definitely wrong. 

For the integration we propose, the most relevant part of the
algorithm comprises L20-31, as it is the code fragment which is
\emph{directly} executed from our abstract interpreter. The remaining
procedures (L1-L19)
will be overridden by more accurate ones later.  
The procedure of  interest
is {\sc specialized\_definition}. As it is customary, it performs
(L21) a generalization of the call $A:CP$ using the abstract
counterpart of the $\abstraction$ operator, denoted by
$\aabstraction$, and which is in charge of ensuring termination at the
global level.
The result of the generalization, $A':CP'$, is inserted in the
generalization table $\gt$ (L22). Correctness of the algorithm
requires that $A:CP\sqsubseteq A':CP'$. If $A':CP'$ has been
previously treated (L23), then its specialized definition $A''$ is
looked up in $\st$ (L24) and returned.  Otherwise, a specialized
definition $\mathit{Def}$ is computed for it by using the $\aunfold$
operator of Def.~\ref{def:aunfold} (L26).
As already mentioned, the specialized definition $\mathit{Def}$ for
the abstract atom $A:\CP{}$ is used to extend the original program
$P$.  First, the atom $A'$ is renamed by using $\newpred$ which
returns an atom with a fresh predicate name, $A''$, and optionally
filters constants out (L27).  Then, function $\ren$ is applied to
rename the clause heads using atom $A'$ (L29).
 $\ren(A,\{B/B'\})$ returns $\theta(B')$ where
$\theta=mgu(A,B)$.  Finally, the program $P$ is extended with the new,
\emph{renamed} specialized definition, $\mathit{Def}'$.

\begin{example}
  
Three calls to {\sc specialized\_definition} appear
(within an oval box) during the analysis of our running example in
Fig.~\ref{fig:aog} from the following abstract atoms, first $\tt
main(s^3(X),X2):\{X/G,X2/V\}$, then $\tt tw(B,C):\{B/G,C/V\}$ and
finally $\tt f(C,A):\{C/G,C/V\}$. The output of such executions is
used later (with the proper renaming) to produce the resultants in
Ex.~\ref{ex:sp}. 
For instance, the second clause obtained from the first call to {\sc
  specialized\_definition} is
\begin{quote}\tt
sp\_main${}_2$(s(s(s(s(B)))),A) :- tw${}_{2,1}$(B,C),formula${}_{2,2}$(C,A).
\end{quote}

\noindent
where only the head is renamed.  The renaming of the body literals is
done in a later code generation phase
%
As already mentioned, Alg.~\ref{algo:pd-with-spec-defs} is not able
to obtain such abstract atoms due to the absence of success
propagation.
\short{
The following section introduces the key concept of how these abstract
atoms are obtained by the abstract interpreter.  Also, we will discuss
later the process of obtaining the partially evaluated program from
the entries in the specialization, in the generalization tables and
the extended program $P'$.
In particular, it is crucial the definition of $\aabstraction$
operators which allow obtaining \emph{feasible} partial evaluations.
}
\end{example}

\short{
to ensure independence. This will also make it possible to generate a
partially evaluated program, as we discuss in the remaining of the
section.
} 

\short{
\begin{theorem}[closedness]
  All calls which might occur  during the execution of the specialized
  program are covered by some program rule.
\end{theorem}
}

\secbeg\secbeg
\section{Abstract Interpretation with Specialized Definitions}
\secend
 \label{sec:effic-prop-abstr}

We now present our final algorithm for abstract interpretation with
specialized definitions. This algorithm extends both the APD
Algorithm~\ref{algo:pd-with-spec-defs} and the abstract interpretation
algorithms in \cite{inc-fixp-sas-short,incanal-toplas}.
W.r.t.\ Algorithm~\ref{algo:pd-with-spec-defs}, the main improvement
is the addition of success propagation.
Unfortunately, this requires computing a global fixpoint. It is an
important objective for us to be able to compute an accurate fixpoint
in an efficient way.
W.r.t the algorithms in \cite{inc-fixp-sas-short,incanal-toplas}, the
main improvements are the following. (1) It deals directly with
non-normalized programs. This point, which does not seem very relevant
in a pure analysis system, becomes crucial when combined with a
specialization system in order to profit from constants propagated by
unfolding. (2) It incorporates a hardwired efficient graph traversal
strategy which eliminates the need for maintaining priority queues
explicitly~\cite{incanal-toplas}.
(3) The algorithm includes a widening operation for calls,
\widencall, which limits the amount of multivariance in order to keep
finite the number of call patterns analyzed. This is required in order
to be able to use abstract domains which are infinite, such as regular
types.
(4) It also includes a number of simplifications to facilitate
understanding, such as the use of the keyed-table ADT, which we assume
encapsulates proper renaming apart of variables and the application
of renaming transformations when needed.

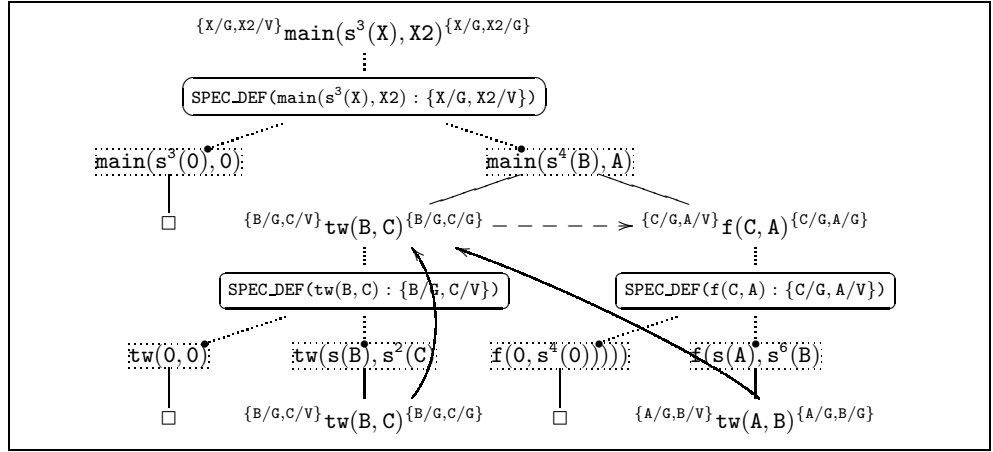
\begin{figure}[t]
\footnotesize \fbox{
\begin{minipage}{12.6cm}\centering
  $\xymatrix@!R=0.5pt{
    &   {\tt ^{\{X/G,X2/V\}}main(s^3(X),X2)^{\{X/G,X2/G\}}} \ar@{..}[d]  \\
    & \ovalbox{\txt{\tt \scriptsize SPEC\_DEF($\tt main(s^3(X),X2):\{X/G,X2/V\}$) }}\ar@{..{*}}[dl]\ar@{..{*}}[dr]\\
   *[F.]{{\tt main(s^3(0),0)}}
    \ar@{-}[d]& & *[F.]{\tt  main(s^4(B),A) } \ar@{-}[dl] \ar@{-}[dr]\\
    \Box & \tt ^{\{B/G,C/V\}}tw(B,C)^{\{B/G,C/G\}}\ar@{..}[d] \ar@{-->}[rr] & &
    \tt ^{\{C/G,A/V\}} f(C,A)^{\{C/G,A/G\}} \ar@{..}[d] \\
& \ovalbox{\txt{\tt  \scriptsize  SPEC\_DEF($\tt tw(B,C):\{B/G,C/V\}$) }}\ar@{..{*}}[dl]\ar@{..{*}}[d] &  &
    \ovalbox{\txt{\tt  \scriptsize SPEC\_DEF($\tt  f(C,A):\{C/G,A/V\}$) }}\ar@{..{*}}[dl]\ar@{..{*}}[d]\\
*[F.]{\tt  tw(0,0)} \ar@{-}[d]  & 
*[F.]{\tt tw(s(B),s^2(C)} \ar@{-}[d]  &  
*[F.]{\tt f(0,s^4(0)))))}  \ar@{-}[d]  &
*[F.]{\tt  f(s(A),s^6(B)} \ar@{-}[d]  \\
 \Box     & \tt ^{\{B/G,C/V\}}tw(B,C)^{\{B/G,C/G\}} \ar@(r,r)[uuu] 
& \Box &  \tt ^{\{A/G,B/V\}}tw(A,B)^{\{A/G,B/G\}}\ar@(r,r)[lluuu]  \\ 
  }$
\end{minipage}
}\caption{Analysis Graph computed by {\scriptsize\sc ABS\_INT\_WITH\_SPEC\_DEF}}
\label{fig:aog}
\end{figure}

In order to compute and propagate success substitutions,
Algorithm~\ref{algo:ai-with-spec-defs} computes a {\em program
  analysis graph} in a similar fashion as state of the art analyzers
such as the \ciaopp\ analyzer~\cite{inc-fixp-sas-short,incanal-toplas}.
%
\short{Such graph can be viewed as a finite representation of the (possibly
infinite) set of (possibly infinite) AND-OR trees explored by the
concrete execution \cite{bruy91}.
} 
For instance, the analysis graph
computed by Algorithm~\ref{algo:ai-with-spec-defs} for our running
example is depicted in Fig.~\ref{fig:aog}.
The graph has two sorts of nodes. Those which correspond to atoms are
called ``OR-nodes''.  For instance, the node $ {\tt
  ^{\{X/G,X2/V\}}main(s^3(X),X2)^{\{X/G,X2/G\}}}$ indicates that when
the atom $\tt main(s^3(X),X2)$ is called with description ${\tt
  \{X/G,X2/V\}}$ the answer (or success) substitution computed is $\tt
\{X/G,X2/G\}$.  Those nodes which correspond to rules are called
``AND-nodes''. In Fig.~\ref{fig:aog}, they appear within a dashed box
and contain the head of the corresponding clause. Each AND-node has as
children as many OR-nodes as literals there are in the body.  If a
child OR-node is already in the tree, it is no further expanded and
the currently available answer is used.
For instance, the analysis graph in Figure~\ref{fig:aog} contains
three occurrences of the abstract atom $\tt tw(B,C):\{B/G,C/V\}$
(modulo renaming), but only one of them has been expanded. This is
depicted by arrows from the two non-expanded occurrences of $\tt
tw(B,C):\{B/G,C/V\}$ to the expanded one.  More information on the
efficient construction of the analysis graph can be found
in~\cite{inc-fixp-sas-short,incanal-toplas,bruy91}.

\short{
\begin{example}
As a result of analyzing our running program, we obtain the following
tables:
\begin{center}
    \begin{tabular}{|c|c|}\hline\hline
     \multicolumn{2}{|c|}{\bf Answer Table} \\ \hline 
     Abstract atom & Answer pattern               \\ \hline\hline
    $\tt \tuple{main(s^3(X),X2)}{\{X/G,X2/V\}}$ & $\tt \{X/G,X2/G\}$  \\ \hline
    $\tt \tuple{tw(B,C)}{\{B/G,C/V\}}$ & $\tt \{B/G,C/G\} $ \\ \hline
$\tt \tuple{f(C,A)}{\{C/G,A/V\}}$ & $\tt \{C/G,A/G\}$ \\\hline
    \end{tabular}
  \end{center}

\begin{center}
\small
    \begin{tabular}{|l|c|}\hline\hline
    \multicolumn{2}{|c|}{\bf Dependency Table} \\ \hline
\textbf{Or-Node} & \textbf{Set of Dependencies}\\ \hline
 $\tt tw(B,C):\{B/G,C/V\}$ & $\tt \{ main(s^3(X),X2):\{X/G,X2/V\} \Rightarrow [main(s^4(B),A):\{B/G,A/V,C/V\}]  2,1 \}$  \\ \hline
$\tt f(C,A):\{A/V,C/G\}$ & $\tt  \{ main(s^3(X),X2):\{X/G,X2/V\}  \Rightarrow
             [main(s^4(B),A):\{B/G,A/V,C/G\}] 2,2\} $ \\ \hline
$\tt tw(D,E):\{D/G,E/V\}$& $\tt \{\{tw(B,C):\{B/G,C/V\} \Rightarrow [tw(s(B),s^2(C):\{B/G,C/V\}]  3,1\},$  \\
       \hline
& \{$\tt \tt  f(D,E):\{D/G,E/V\} \Rightarrow [f(s(B),s^6(C)):\{B/G,C/V\}]  5,1\}\}$  \\\hline

    \end{tabular}
  \end{center}

Note that there is an entry in the answer table for each abstract atom which
appear in the graph. The dependencies correspond to
\mycomment{complete}. The specialization table contains
\mycomment{complete}. Finally, the generalization table is empty
because \mycomment{complete}. 

\end{example}
 
} 

\short{
Finiteness of the program analysis
graph (and thus termination of the algorithm) is achieved by
considering description domains with certain characteristics (such as
being finite, or of finite height, or without infinite ascending
chains) or by the use of a {\em widening} operator \cite{Cousot77-short}.
The graph has two sorts of nodes: those belonging to clauses (also
called ``AND-nodes'') and those belonging to atoms (also called
``OR-nodes'').  
Atoms in the clause body have arcs to OR-nodes with the corresponding
calling pattern. If such a node is already in the graph it becomes a
recursive call.
} 

\begin{algorithm}[t]
\caption{Abstract Interpretation with Specialized Definitions}
\label{algo:ai-with-spec-defs}

\begin{algorithmic} [1]
\renewcommand{\baselinestretch}{0.5} \sizeinfigure
\Procedure{abs\_int\_with\_spec\_defs}{$P,\{A_1:CP_1,\ldots,A_n:CP_n\}$}
\State \emptytable$(\at)$; \emptytable$(\dt)$
\State \emptytable$(\gt)$; \emptytable$(\st)$
\For{$j = 1..n$}
\State  {\sc process\_call\_pattern}($A_j:CP_j,\dep{A_j:CP_j\Rightarrow{}[A_j:CP_j],j,entry}$) 
\EndFor
\EndProcedure
\Statex
\Function{process\_call\_pattern}{$A:CP,Parent$}
\State $CP_1 \gets \widencall(\at,A:CP)$
\If {{\bf not }\isin$(\at,A:CP_1)$}
\State \store$(\at,A:CP_1,\bot)$
\State \store$(\dt,A:CP_1,\emptyset)$
\State $(A',A_1')\gets$ {\sc specialized\_definition}$(P,A:CP_1)$
\State $A''\gets \ren(A,\{A'/A_1'\})$
\ForAll  {ren. apart clause $C_k=H_k \leftarrow B_{k}\in P$
  s.t. $H_k$ unifies with $A''$}
\State $CP_k$ $\gets$ \acalltoentry$(A'':CP_1,C_k)$
\State {\sc process\_clause}$(A:CP_1 \Rightarrow{} [H_k:CP_k]~B_{k},k,1)$
\EndFor
\EndIf
\State $Deps\gets \consult(\dt,A:CP_1)\bigcup \{Parent\}$ 
\State \store$(\dt,A:CP_1,Deps$)
\State \Return $\consult(\at,A:CP_1)$
\EndFunction
\Statex
\Procedure{process\_clause}{$H:\!CP \Rightarrow  [H_k:CP_1]~B,k,i$}
\If {$CP_1 \neq \bot$}
\If {$B=(L,R)$}
\State $CP_2 \gets \arestrict(CP_1,L)$
\State $AP_0$ $\gets$ {\sc process\_call\_pattern}($L:CP_2,\dep{H:\!CP \Rightarrow  [H_k:CP_1],k,i}$)  
\State $CP_3$ $\gets$ \aconj$(CP_1,\aextend(AP_0,CP_1))$ 
\State {\sc process\_clause}$(H:CP \Rightarrow [H_k:CP_3]~R,k,i+1)$  
\Else 
\State $CP_2 \gets \arestrict(CP_1,B)$
\State $AP_0\gets$ {\sc process\_call\_pattern}($B:CP_2,\dep{H:CP \Rightarrow [H_k:CP_1],k,i}$)  
\State $CP_3$ $\gets$ \aconj$(CP_1,\aextend(AP_0,CP_1))$ 
\State $AP_1$ $\gets$ \acalltoentry$(H_k:CP_3, H\leftarrow true)$ 
\State $AP_2 \gets$ \consult$(\at,H:CP)$
\State $AP_3 \gets$ \alub$(AP_1,AP_2)$ 
\If {$AP_2 \neq AP_3$} 
\State \store$(\at,H:CP,AP_3)$ 
\State $Deps\gets$ \consult$(\dt,H:CP)$ 
\State {\sc process\_update}($Deps$)
\EndIf
\EndIf
\EndIf
\EndProcedure
\Statex
\Procedure{process\_update}{$Updates$}
\If {$Updates = \{A_1,\ldots,A_n\}$ with $n\geq 0$}
\State $A_1=\dep{H:\!CP \Rightarrow  [H_k:CP_1],k,i}$
\If {$i\neq entry$}
\State $B\gets \getbody(P,k,i)$
\State {\sc remove\_previous\_deps}($H:\!CP \Rightarrow  [H_k:CP_1]~B,k,i$)
\State {\sc process\_clause}$(H:\!CP \Rightarrow  [H_k:CP_1]~B,k,i)$
\State {\sc process\_update}($Updates - \{A_1\}$)
\EndIf
\EndIf
\EndProcedure
\renewcommand{\baselinestretch}{1.0}\normalsize
\end{algorithmic}
\end{algorithm}

The program analysis graph is implicitly represented in the algorithm
by means of two data structures, the {\em answer table} (\at) and the
{\em dependency table} (\dt).
The answer table contains entries of the form $A:\CP{} \leadsto
\AP{}$ which  
are interpreted as the answer (success) pattern for $A:\CP{}$ is
$\AP{}$.  For instance, there exists an entry of the form $\tt
main(s^3(X),X2):\{X/G,X2/V\} \leadsto \{X/G,X2/G\}$ associated to the
atom discussed above.  Dependencies indicate direct relations among
OR-nodes. An OR-node $A_F:CP_F$ \emph{depends on} another OR-node
$A_T:CP_T$ iff in the body of some clause for $A_F:CP_F$ there appears
the OR-node $A_T:CP_T$. The intuition is that in computing the answer
for $A_F:CP_F$ we have used the answer pattern for $A_T:CP_T$. In our
algorithm we store \emph{backwards} dependencies,\footnote{In the
  implementation, for efficiency, both forward and backward
  dependencies are stored. We do not include them in the algorithm for
  simplicity of the presentation.} i.e., for each OR-node $A_T:CP_T$
we keep track of the set of OR-nodes which depend on it. That is to
say, the keys in the dependency table are OR-nodes and the information
associated to each node is the set of other nodes which depend on it,
together with some additional information required to iterate when an
answer is modified (updated).  Each element of a \emph{dependency set}
for an atom $B:CP_2$ is of the form $\dep{H:\CP{} \Rightarrow
  [H_k:\CP{}_1] ~k,i}$.
It should be interpreted as follows: the OR-node $H:\CP{}$ through the
literal at position $k,i$ depends on the OR-node $B:CP_2$. Also, the
remaining information $[H_k:\CP{}_1]$ informs that the head of this
clause is $H_k$ and the substitution (in terms of all variables of
clause $k$) just before the call to $B:CP_2$ is $CP_1$.  Such
information avoids reprocessing atoms in the clause $k$ to the left of
position $i$. For instance, the dependency set for $\tt
f(C,A):\{A/V,C/G\}$ is $\tt \{ \dep{main(s^3(X),X2):\{X/G,X2/V\}
\Rightarrow [main(s^4(B),A):\{B/G,A/V,C/G\}] 2,2}\}$. It indicates that
the OR-node $\tt f(C,A):\{A/V,C/G\}$ is only used, via literal (2,2), in the OR-node $\tt
main(s^3(X),X2):\{X/G,X2/V\}$  (see
Example~\ref{ex:sp}). Thus, if the answer pattern for $\tt
f(C,A):\{A/V,C/G\}$ is ever updated, then we must reprocess the
OR-node $\tt \{ main(s^3(X),X2):\{X/G,X2/V\}$ from position 2,2.

\short{
Intuitively, the analysis algorithm is just a graph traversal
algorithm which places entries in the answer table and dependency
table as new nodes and arcs in the program analysis graph are
encountered.
} 

Algorithm~\ref{algo:ai-with-spec-defs} proceeds as follows. The
procedure {\sc abs\_int\_with\_spec\_defs} initializes the four tables
used by the algorithm and calls {\sc process\_call\_pattern} for each
abstract atom in the initial set.
The function {\sc process\_call\_pattern} applies, first of all (L7),
the \widencall\ function to $A:CP$ taking into account the set of
entries already in \at. This returns a substitution $CP_1$ s.t. $CP
\sqsubseteq CP_1$. The most precise \widencall\ function possible is
the identity function, but it can only be used with abstract domains
with a finite number of abstract values. This is the case with
\emph{sharing--freeness} and thus we will use the identity function in
our example.
If the call pattern $A:CP_1$ has not been processed before, it places
(L9) $\bot$ as initial answer in \at\/ for $A:CP$ and sets to empty
(L10) the set of OR-nodes in the graph which depend on $A:CP_1$. It
then computes (L11) a specialized definition for $A:CP_1$.
We do not show in Algorithm~\ref{algo:ai-with-spec-defs} the
definition of {\sc specialized\_definition}, since it is identical to
that in Algorithm~\ref{algo:pd-with-spec-defs}. In the graph, we show
within an oval box the calls to {\sc specialized\_definition} which
appear during the execution of the running example (see the details in
Sect.~\ref{sec:spec-defin}). The clauses in the specialized definition
are linked to the box with a dotted arc.  Then it launches (L13-15)
calls to {\sc process\_clause} for the clauses in the specialized
definition w.r.t. which $A:CP_1$ is to be analyzed. Only after this,
the \emph{Parent} OR-node is added (L16-17) to the dependency set for
$A:CP_1$.

The function {\sc process\_clause} performs the success propagation
and constitutes the core of the analysis.  
First, the current answer ($\AP{}_0$) for the call to the literal at
position $k,i$ of the form $B:\CP{}_2$ is (L24 and L29) conjoined
(\aconj), after being extended (\aextend) to all variables in the
clause, with the description $\CP{}_1$ from the program point
immediately before $B$ in order to obtain the description $\CP{}_3$
for the program point after $B$. If $B$ is not the last literal,
$CP_3$ is taken as the (improved) calling pattern to process the next
literal in the clause in the recursive call (L25). This corresponds to
left-to-right success propagation and is marked in Fig.~\ref{fig:aog}
with a dashed horizontal arrow.  If we are actually processing the
last literal, $\CP{}_3$ is (L30) adapted (\acalltoentry) to the
initial call pattern $H:CP$ which started {\sc process\_clause},
obtaining $\AP{}_1$.  This value is (L32) disjoined (\alub) with the
current answer, $\AP{}_2$, for $H:CP$ as given by {\consult}.  If the
answer changes, then its dependencies, which are readily available in
\dt, need to be recomputed (L36) using {\sc process\_update}. This
procedure restarts the processing of all body postfixes which depend
on the calling pattern for which the answer has been updated by
launching new calls to {\sc process\_clause}.  There is no need of
recomputing answers in our example.  The procedure {\sc
  remove\_previous\_deps} eliminates (L42) entries in \dt\/ for the clause
postfix which is about to be re-computed. We do not present its
definition here due to lack of space. Note that the new calls to {\sc
  process\_clause} may in turn launch calls to {\sc process\_update}.
On termination of the algorithm a global fixpoint is guaranteed to
have been reached. Note that our algorithm also stores in the
dependency sets calls from the initial entry points (marked with the
value $\mathit{entry}$ in L5). These do not need to be reprocessed
(L40) but are useful for determining the specialized version to use
for the initial queries after code generation.

\secbeg
\subsection{Termination of Abstract Interpretation with Specialized Definitions}
\secend

Termination of Algorithm~\ref{algo:ai-with-spec-defs} comprises
several levels. First, termination of the algorithm requires the local
termination of the process of obtaining a specialized definition. This
corresponds to ensuring termination of function {\sc
  specialized\_definition} in Algorithm~\ref{algo:pd-with-spec-defs}.
Second, we need to guarantee that the number of call patterns for
which a specialized definition is computed is finite. This corresponds
to global termination of specialization algorithms.  In terms of our
algorithm, this is equivalent to having a finite number of entries in
\st. The \aabstraction\ function should be able to guarantee it.
Third, it is required that the set of call patterns for which an
answer pattern is to be computed be finite. This corresponds to
control of multivariance in context-sensitive analysis. In terms of
our algorithm, this is equivalent to having a finite number of entries
in \at. The \widencall\ function should be able to guarantee it.
Fourth and final, it is required that the computation of the answer
pattern for each entry in \at\/ needs a finite number of iterations.
This is guaranteed since we consider domains which are ascending chain
finite.
%
%
Another way of looking at this problem is that, intuitively, the
combined effect of terminating \aunfold\ and \aabstraction\ operators
guarantee that the set of specialized definitions which
Algorithm~\ref{algo:ai-with-spec-defs} will compute for an initial set
of atoms is finite. These two problems have received considerable
attention by the PD community (see, e.g.,
\cite{LeuschelBruynooghe:TPLP02}).
Since
Algorithm~\ref{algo:ai-with-spec-defs} performs analysis of the program
composed of the set of specialized definitions, once we have
guaranteed the finiteness of the program to be analyzed, a terminating
\widencall\ together with an abstract domain which is ascending chain
finite guarantee termination of the whole process.
\comment{GP: maybe there is no room for this other explanation
Another way of looking at this problem is that if we have a
terminating \aunfold\ rule and the abstract domain is ascending chain
finite, non-termination can only occur if the set of call patterns
handled by the algorithm is infinite. 
Since the \widencall\ function guarantees that a given concrete atom
$A$ can only be analyzed w.r.t. a finite number of abstract
substitutions $CP$, non-termination can only occur if the set the set
of atoms has infinite elements with different concrete parts. 
If the \aabstraction\ function guarantees that an infinite number of
different concrete atoms cannot occur, then non-termination cannot
occur.
}
 
 

\secbeg
\section{Discussion and Related Work}
\secend
 \label{sec:disc-relat-work}

We have presented a generic framework for the analysis and
specialization of logic programs which is currently the basis of the
analysis/specialization system implemented in the \ciaopp\ 
preprocessor. We argue that, in contrast to other approaches, the fact
that our method can be used both as a specializer and analyzer gives
us more accuracy and efficiency than the individual techniques.
Indeed, the versatility of our framework (and of our implementation)
can be seen by recasting well-known specialization and analysis
frameworks as instances in which the different parameters:
unfolding rule, 
widen call rule,
abstraction operator,
 and
analysis domain, take the following values.

\secbeg
\paragraph{Polyvariant Abstract Interpretation:} Our algorithm can behave
as the analysis algorithm described in
\cite{incanal-toplas,inc-fixp-sas-short} for polyvariant static
analysis by defining a $\aabstraction$ operator which returns always
the base form of an expression (i.e., it loses all constants) and an
$\aunfold$ operator which performs a single derivation step (i.e., it
returns the original definition). Thus, the resulting framework would
always produce a residual program which coincides with the original
one and can be analyzed with any abstract domain of interest.
  
\secbeg
\paragraph{Multivariant Abstract Specialization:} The specialization power
of the framework described in \cite{spec-inv-pepm03-short,spec-jlp-short} can be
obtained by using the same $\aabstraction$ described in the above point
plus an $\aunfold$ operator which always performs a derive step
followed by zero or more abstract execution steps.
It is interesting to note that in the original framework abstract
executability is performed as an offline optimization phase while it
is performed online in our framework.
  
\secbeg
\paragraph{Classical Partial Deduction:} Our method can be used to perform
classical PD in the style of
\cite{Lloyd91,gallagher:pepm93-short} by using an abstract domain with the
single abstract value $\top$ and the identity function as \widencall\ 
rule. This corresponds to the ${\cal PD}$ domain
of~\cite{leuschel04:Toplas} in which an atom with variables represents
all its instances.  Let us note that, in spite of the fact that the
algorithm follows a left-to-right computation flow, the process of
generating specialized definitions (as discussed in
Section~\ref{sec:unfold-with-abstr}) can perform \emph{non-leftmost}
unfolding steps and achieve optimizations as powerful as in PD.
  
\secbeg
\paragraph{Abstract Partial Deduction:}  Several approaches  have been
proposed which extend PD by using abstract
substitutions~\cite{leuschel-jicslp98,Gallagher-Peralta-HOSC,leuschel-gruner-loprstr2001-short,leuschel04:Toplas}.
In essence, such approaches are very similar to the abstract partial
deduction with call propagation shown in
Algorithm~\ref{algo:pd-with-spec-defs}. Though all those proposals
identify the need of propagating success substitutions, they either
fail to do so or propose means for propagating success information
which are not fully integrated with the APD algorithm and, in our
opinion, do not fit in as nicely as the use of and--or trees. Also,
these proposals are either strongly coupled to a particular (downward
closed) abstract domain, i.e., regular types, as in
\cite{Gallagher-Peralta-HOSC,leuschel-gruner-loprstr2001-short} or do
not provide the exact description of operations on the abstract domain
which are needed by the framework, other than general correctness
criteria~\cite{leuschel-jicslp98,leuschel04:Toplas}. However, the
latter allow conjunctive PD, which is not available in our framework.

\secbeg
\paragraph{The approach in~\cite{pe-in-plai-pepm99-short}:} was a starting step
  towards our current framework.
  There, the introduction of unfolding steps directly 
  in the and--or graph was proposed in order to achieve
  transformations as powerful 
  as those of PD while at the same time propagating abstract
  information.
  In contrast, we now resort to augmented SLD semantics for the
  specialization side of the framework while using AND-OR semantics
  for the analysis side of the framework.  This has both conceptual,
  the hybrid approach we propose provides satisfactory answers to the
  four issues raised in Section~\ref{sec:motivation}, and practical
  advantages, since the important body of work in control of PD is
  directly applicable to the specialization side of our framework.

\secbeg
\section{Conclusions}
\secend
 \label{sec:benefits-or-our}

We have proposed a novel scheme for a seamless integration of the
techniques of abstract interpretation and partial deduction.  Our
scheme is parametric w.r.t.\ the abstract domain and the control
issues which guide the partial deduction process.  Existing proposals
for the integration use abstract interpretation as a \emph{means} for
improving partial evaluation rather than as a \emph{goal}, at the same
level as producing a specialized program.  This implies that, as a
result, their objective is to yield a set of atoms which determines a
partial evaluation rather than to compute a safe approximation of its
success.  Unlike them, a main objective of our work is to improve
success information by analyzing the specialized code, rather than the
original one. We achieve this objective by smoothly
\emph{interleaving} both techniques which improves
success information---even for abstract domains which are not related
directly to partial evaluation. Moreover, with more accurate
success information, we can improve further the quality of partial
evaluation.  The overall method thus yields not only a specialized
program but also a safe approximation of its behaviour.

\comment{

Our new framework conceptually distinguishes two interleaved phases:
one in which specialized definitions are computed and a second one in
which success substitutions are obtained by traversing such
specialized definitions from left to right until a global fixpoint is
reached.

In some examples, we observed that by applying \emph{several}
consecutive times a sequence of the above individual techniques we can
achieve the same effect than by applying \emph{once} our algorithm.
The intuition is that after unfolding we can improve the analysis
information which at the same time will improve unfolding in the next
iteration. There are two lines of ongoing research related to this
issue. First, we are currently studying whether the number of
iterations that the individual techniques have to be applied is
actually bounded. Then, we also want to compare the efficiency of our
technique with the application consecutive of the individual
techniques.

The benefits of our approach can be seen from two perspectives, from
the point of view of analysis and from that of program
specialization. 

Though our approach can be thought of as a powerful specialization
method, it also brings quite important benefits to program
analysis. The success substitutions computed by our algorithm by using
specialized definitions instead of the original ones, not only is
guaranteed to achieve correct results, but also will in general
achieve more accurate results. The alternative to computing
specialized definitions in order to achieve success substitutions of
comparable accuracy is to use abstract domains which capture precise
values of variables, terms, etc. A possible choice is the use of
abstract domain based on regular types. Since our algorithm is
context-sensitive and polyvariant, such domains require the use of a
non-trivial polyvariance control, in our algorithm represented by the
\widencall\ function, in order to guarantee termination since the
number of different call patterns for a same predicate is potentially
infinite. 

When the program or the initial descriptions contain relatively large
amounts of \emph{concrete} data, analysis using regular types either
loses precision very quickly in order to guarantee efficiency (for
example, by being monovariant on calls) or become impractical. Another
difficulty related to the use of abstract interpretation only is that
there is very quickly no way to distinguish whether abstract
substitution originate from concrete data in the program and/or
initial descriptions or from previous computation of success
substitutions. In our experience this distinction is important since
concrete data should be consumed during computation of specialized
definitions, whereas abstract information can be treated more
conservatively in order to keep the cost of analysis within reasonable
bounds. 

From the point of view of specialization, the possibility of having
context-sensitive success information is very important.  It fits very
nicely in the framework of call propagation. Moreover, it may be the
case that context-sensitive analysis may provide a higher degree of
accuracy which may be crucial in order to enable certain
specializations.

}

\vspace{-0.5cm} 
\secbeg
\subsection*{Acknowledgments}
\secend

\begin{small}
The authors would like to thank John Gallagher and Michael Leuschel
for useful discussions on the integration of abstract interpretation
and partial deduction.
 This work was
  funded in part by the Information Society Technologies programme of
  the European Commission, Future and Emerging Technologies under the
  IST-2001-38059 {\em ASAP} project and by the Spanish Ministry of
  Science and Education under the MCYT TIC 2002-0055 {\em CUBICO}
  project.
  Manuel Hermenegildo is also supported by the Prince of Asturias
  Chair in Information Science and Technology at UNM.2
\end{small}
\vspace{-0.5cm} 
\secbeg
\secbeg





\end{document}